# Computer Network Topology Design in Limelight of Pascal Graph Property

### Sanjay Kumar Pal[1] and Samar Sen Sarma[2]

[1]Department of Computer Science & Applications, NSHM College of Management & Technology, Kolkata, INDIA,
pal.sanjaykumar@gmail.com

[2]Department of Computer Science & Engineering University of Calcutta, Kolkata, INDIA
sssarma2001@yahoo.com

## Abstract

*Constantly growing demands of high productivity and security of computer systems and computer networks call the interest of specialists in the environment of construction of optimum topologies of computer mediums. In earliest phases of design, the study of the topological influence of the processes that happen in computer systems and computer networks allows to obtain useful information which possesses a significant value in the subsequent design. It has always been tried to represent the different computer network topologies using appropriate graph models. Graphs have huge contributions towards the performance improvement factor of a network. Some major contributors are de-Bruijn, Hypercube, Mesh and Pascal. They had been studied a lot and different new features were always a part of research outcome. As per the definition of interconnection network it is equivalent that a suitable graph can represent the physical and logical layout very efficiently. In this present study Pascal graph is researched again and a new characteristics has been discovered. From the perspective of network topologies Pascal graph and its properties were first studied more than two decades back. Since then, a numerous graph models have emerged with potentials to be used as network topologies. This new property is guaranteed to make an everlasting mark towards the reliability of this graph to be used as a substantial contributor as a computer network topology. This shows its credentials over so many other topologies. This study reviews the characteristics of the Pascal graph and the new property is established using appropriate algorithm and the results.*

## 1. Introduction

While searching for a class of graphs with certain desired properties to be used as computer networks, we have found graphs that come close to being optimal. One of the desired properties is that the design be simple and recursive, so that when a new node is added, the entire network does not have to be reconfigured. Another property is that one central vertex is adjacent to all others. The third requirement is that there exists several paths between each pair of vertices (for reliability) and that some of this paths be of short lengths, to reduce communication delay. Finally the graphs should have good cohesive and connectivity. Complete graph $K_n$ satisfy all these properties, but are ruled out because of the expense.

The proper introduces a set of adjacency matrices called Pascal matrices, which are constructed using Pascal's triangle modulo 2. We also define Pascal graphs, the set of graphs corresponding to Corresponding to Pascal matrices.

In this paper, author tries in summarized way, to expose the fundamental problems of topological design in computer technology [8, 9, 10, 11, 12]. The problems of multiprocessing and computer systems have been formulated, and these problems have been resolved in form of classes of Pascal graph with extreme topological parameters that correspond to the parameters of speed of information transmission, productivity, and security with limitations in costs [10]. In the phase of topological design, the computer network is represented in graph [8, 10] form whose vertices correspond to the nodes of information processing, and edges correspond to the communication lines.

The ontogeny of Pascal Graph (PG) was Pascal Matrix (PM) that in turn was generated meticulously from Pascal's triangle [1]. Scientist have been putting a lot of efforts in ameliorating computer network properties [4]. Wide varieties of graph model worked as resource to their brainstorming contribution in this field [3]. Pascal graph is one of those resources and played a significant role as soon as it exploration was initiated almost two decades back. As a consequence, several similar graph models emerged in this arena with laudable potentials to be used as computer network topologies [2]. The inspiration of this study stemmed from consistent urge to contribute substantially in this crescendo of research from the perspective of computer network topology. Here we have reincarnated Pascal Graph with another feather added to its credentials. We have reviewed to its characteristics and established a significant property with adequate theoretical practical support. The technique and algorithms here proposed can be used not only in the design of computer networks, but also in the design of other network.

## 2. Origin of Pascal Graph

We introduced Pascal Graph from historical modern perspective. Blaise Pascal (1623 - 1662) first conceptualized Pascal's Triangle around the middle of seventeenth century [7]. This Pascal triangle played the most important role while generating Pascal Matrix [1].

### 2.1 Pascal Matrix

An (n x n) symmetric binary matrix is called the Pascal Matrix PM(n) of order n if its main diagonal entries are all 0's and its lower (and therefore the upper also) consists of the first (n-1) rows of Pascal Triangle modulo 2. Where $pm_{i,j}$ denotes the element of $i^{th}$ row and $j^{th}$ column of the Pascal Matrix [1]. An example of Pascal graphs along with associated Pascal matrices is shown in next section.

### 2.2 Pascal Graph

An undirected graph of n vertices corresponding to PM(n) as an adjacency matrix is called Pascal Graph (n), where n is the order of the Pascal graph [1]. An example of Pascal graph having 5 nodes is shown hereunder, for better comprehension of the study.

Pascal Graph (5),

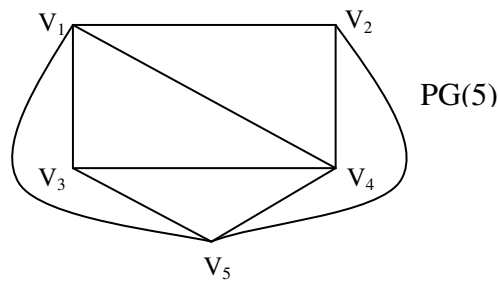

Pascal Matrix(5),

|       | $V_1$ | $V_2$ | $V_3$ | $V_4$ | $V_5$ |       |
|-------|-------|-------|-------|-------|-------|-------|
| $V_1$ | 0     | 1     | 1     | 1     | 1     |       |
| $V_2$ | 1     | 0     | 1     | 0     | 1     |       |
| $V_3$ | 1     | 1     | 0     | 1     | 1     | PM(5) |
| $V_4$ | 1     | 0     | 1     | 0     | 1     |       |
| $V_5$ | 1     | 1     | 1     | 1     | 0     |       |

## 2. 3 Connectivity Properties of Pascal Graph

There are certain pragmatic properties that make Pascal graph a better choice for a computer network topology over many others. Some of those properties are given below:

- PG(n) is a subgraph of PG(n+1) $\forall \, n \geq 1$.
- All Pascal Graph PG(i) for $i \leq 1 \leq 7$ are planner; all Pascal Graph of higher order are non-planner.
- Vertex $V_1$ is adjacent to all other vertices in the Pascal Graph. Vertex $V_1$ is adjacent to $V_{i+1}$ in the Pascal graph for $i \geq 1$.
- PG(n) contains a star tree $\forall \, n \geq 1$.
- PG(n) contains a Hamiltonian circuit [1,2,3,……,n-1, n,1].
- PG(n) contains $w_{n-x}$ (wheel of order n minus an edge).
- If $k = 2^n + 1$, n is a positive integer, then $V_k$ is adjacent to all $V_i$.
- All Pascal Graph of order $\geq 3$ are 2-connected.
- No two even no of vertices of a Pascal Graph are adjacent.
- There are at least two edge disjoint path of length $\leq 2$ between any two distinct vertices in PG(n), $3 \leq n$.
- If $V_i$ is adjacent to $V_j$, where j is even and |i-j|>1, then I is odd and $V_i$ is adjacent to $V_{j-1}$.
- Let det(PM(n)) refer to the determinant of the Pascal matrix of order n. Then, det(PM(n)) = 0, for all even $n \geq 4$.
- Define e(PG(n)) to be the number of edges in PG(n), $c(PG(n)) \leq \lfloor (n-1)\log_2 3 \rfloor$.
- Det(PM(n)) is even for all $n \geq 3$.

## 2.4 Pascal Graph Generation Algorithm

***Step 1:*** Enter n number of vertices.

***Step 2:*** Initialized LT[n,0] = 1;

***Step 3:*** From the lower left triangle [LT(n,n)] by adding the number directly above and to the left with the number directly above and to th right to find the new value. If either the number to the right or left is not present, substitute a zero to its place.

***Step 4:*** From the upper right triangle [UT(n,n)] using the same manner.

***Step 5:*** Convert the lower left triangle into binary values.

***Step 6:*** Convert the upper right triangle into binary values.

***Step 7:*** From the final adjacency matrix [PM(n,n)] of with LT(n,n) and UT(n,n).

***Step 8:*** Stop.

## 3. Connotation of Present Work

It is better not to predestine any claim before underpinned with adequate reasoning. To bolster this, the basic requirement must be articulated and hence substantiated. Studying different graph with respect to their properties has been a basic research interest for a long time [6]. It can be said without any ambiguity that there are certain criteria in any network topology that are given priorities to judge its credentials. Those criteria are, (i) degree of the network, (ii) diameter of the network, (iii) scalability of the network and (iv) reliability of the network [2]. There is a tradeoff between degree related to hardware cost and diameter related to transmission time of messages within any Interconnection Network [2]. Reliability is the quality to sustain faulty circumstances and scalability is the scope of up-gradation without much alteration. To prove its worth, Pascal; graph have been merged cogently with Hypercube graph to engender simple-(q,n)-graph [2]. The destiny of the research is signified keeping this philosophy in mind.

## 4. Fascinating Characteristic of Pascal Graph

In this work a new property of Pascal graph has been characterized that has increased the reliability of the graph without perturbing its degree, diameter as well as scalability as computer network topology. As we mentioned in property (iii) in properties of Pascal graph, vertex $v_1$ of any Pascal graph is adjacent to all other vertices in that graph. It has been shown that if the index start from 0 instead of 1 then it can be proved that vertex $v_0$ in any Pascal graph is adjacent to all other vertices in that graph. Vertex $v_1$ is adjacent to $v_{i+1}$ in the Pascal graph if $i \geq 0$ [2]. Extending the property (iii) in the properties of Pascal graph, it can be very well shown that there exist other node $v_i$ ( $i \geq 0$ ) having the same property like $v_1$. According to property (iv), this node will never have even number as its index.

### 4.1 Thought of Present Work

If we consider a Pascal graph containing n nodes i.e. PG(n) then by default $v_i$ ( $i \geq 1$ ) is adjacent to all other nodes [1]. The exploration does not conclude with it. We have found and recommended other node similar to $v_1$ in the PG(n) and termed it as Dependable Node of Pascal Graph (DNP); the special node, other than node $v_1$, of Pascal graph PG(n) with same degree like $v_1$.

To satiate our claim and to rationalize the whole thing we have used very simplistic approach. The explanation with some suitable example is given below.

Case 1: $n = 2^{(\log n)}$

Case 2: $n \, \pi \, 2^{\lceil \log n \rceil}$

Case N: $n = 2^{\lfloor \log n \rfloor} + 1$

To establish the new property, we are supposed to generate the index I of DNP. Here, we provide formulae for each case including the case 3.

For Case 1: $i = 2^{(\log n)-1} + 1$, *where i* $\phi$ 1

For Case 2: $i = 2^{\lfloor \log n \rfloor} + 1$ *where i* $\phi$ 1

For Case N: $i = 2^{\lfloor \log n \rfloor - 1} + 1$ *and* $i = 2^{\lfloor \log n \rfloor} + 1$, *where i* $\phi$ 1

### 4.2 Experimental Result

The source code has been written in the 'C' programming language to generate the different order Pascal matrix. The formulas were put into test for a meticulous checking of the existence of Dependable Node Pascal Graph (DNP) in different Pascal matrices generated by the program. Some instances of execution of the program are given in the table 1, for a better comprehension of the conjecture. Each instances given in the table 1, the Pascal matrix has been generated and printed along with the degree of the particular node and its index. When case N occurs, the Pascal graph is at at a precise level of reliability. Unlike other cases, it has got three (one more than case 1 and case 2) nodes.

| Table 1: | | | | |
|---|---|---|---|---|
| PG(n) | Conjecture satisfied | i | DNP | Degree |
| 8 | Case 1 | 5 | $V_5$ | 7 |
| 9 | Case N | 5, 9 | $V_5$ , $V_9$ | 8 |
| 10 | Case 1 | 9 | $V_9$ | 9 |
| 11 | Case 2 | 9 | $V_9$ | 10 |
| 14 | Case 1 | 9 | $V_9$ | 13 |
| 15 | Case 2 | 9 | $V_9$ | 14 |
| 16 | Case N | 9 | $V_9$ | 15 |
| 17 | Case N | 9, 17 | $V_9$, $V_{17}$ | 16 |
| 32 | Case 1 | 19 | $V_{17}$ | 31 |
| 33 | Case N | 17, 33 | $V_{17}$, $V_{33}$ | 32 |
| 34 | Case 2 | 33 | $V_{33}$ | 33 |
| 40 | Case 2 | 33 | $V_{33}$ | 39 |

## 5. Significance of the Work

When Pascal graph merged with Hypercube to form Simple-(q,n)-graph [2]; it showed great credentials to be exploited as a typical computer network topology. It alone can contribute substantially in this field of active research. This new discovery will earn Pascal graph high esteem through its reliability under erratic and exigent erroneous circumstances and make it sure that the situation does not exacerbate any more. As per property (iii) mentioned in some property of Pascal graph, it is very easy to access or traverse any node from $v_1$ directly in a single hop. Every other node except $v_1$ had to use either two hop (the two communicating nodes have even number indices) or single hop otherwise. Everything is fine till $v_1$ is not perturbed. If, under any faulty circumstances, $v_1$ gets disconnected then the above mentioned property no longer holds for Pascal graph and the reliability is seized with an increase in the minimum number of hops required by the message to travel from one node to another. In this paper we have reestablished its reliability under this circumstances by finding DNP that is similar to $v_1$ in PG(n). by this work we have eradicated threat imposed upon Pascal graph by mollifying the faulty circumstances. The superiority of the graph is maintained very efficiently.

As the whole work is not based on any convoluted formulae or theory, it is easy to exploit Pascal graphs' high reliability, effective degree and diameter from computer network point of view. Now, it is a reliable choice for any network designer to opt the Pascal graph when reliability along with efficient degree and diameter are the key factors in designing. Since the new property is based on very simple logic it will be easier to incorporate the concept into the physical and logical layout of the computer network topology.

## 6. Scope of Future work

Selecting various graph models to represent topologies of computer network had always been an active research area for network Scientists [5]. The reasonable criteria behind this selection are very significant because this put long testing impact on the network. How to make a graph model an ideal choice to implement a specific topology would always impose a huge challenge to the upcoming Scientists. The chance of finding optimal, if not optimum, graph model with unparallel properties entirely depends on the institution of the Scientists, in-depth analysis of the graph and technical feasibility of implementing computer network using that graph. Finally, it can be said very intuitively that this field still holds immense scopes for Scientists for further research work.

## 7. Conclusion

A source code has been written in C programming language to generate the Pascal Matrix of different order. The formulae were put into test for a meticulous checking of the existence of DNP in different Pascal Matrices generated by the program. From the case 3, the Pascal graph is at a precise level of reliability. The Pascal Matrix has been generated along with some extra details e.g. the degree of particular node and its index. Unlike other cases, it has got three nodes that are adjacent to all other nodes.